\documentclass[graybox]{svmult}

\usepackage{mathptmx}       
\usepackage{helvet}         
\usepackage{courier}        
\usepackage{type1cm}        
\usepackage{makeidx}         
\usepackage{graphicx}        
\usepackage{multicol}        
\usepackage[bottom]{footmisc}

\makeindex             

\def\be{\begin{eqnarray}}
\def\ee{\end{eqnarray}}
\def\bdm{\begin{displaymath}}
\def\edm{\end{displaymath}}

\begin{document}

\title*{Modeling space plasma dynamics with anisotropic Kappa distributions}
\author{M. Lazar, V. Pierrard, S. Poedts and R. Schlickeiser}
\institute{M. Lazar \at Institut f\"ur Theoretische Physik,
Lehrstuhl IV: Weltraum- und Astrophysik, Ruhr-Universit\"at Bochum,
D-44780 Bochum, Germany, \email{mlazar@tp4.rub.de} \and V. Pierrard
\at Belgian Institute for Space Aeronomy, Space Physics, av.
circulaire 3, 1180 Brussels, Belgium \email{viviane.pierrard@oma.be}
\and S. Poedts \at Centre for Plasma Astrophysics, Celestijnenlaan
200B, 3001 Leuven, Belgium, \email{Stefaan.Poedts@wis.kuleuven.be}
\and R. Schlickeiser \at Institut f\"ur Theoretische Physik,
Lehrstuhl IV: Weltraum- und Astrophysik, Ruhr-Universit\"at Bochum,
D-44780 Bochum, Germany, \email{rsch@tp4.rub.de}}
%
%
\maketitle

\abstract{Space plasmas are collisionpoor and kinetic effects
prevail leading to wave fluctuations, which transfer the energy to
small scales: wave-particle interactions replace collisions and
enhance dispersive effects heating particles and producing
suprathermal populations observed at any heliospheric distance in
the solar wind. At large distances collisions are not efficient, and
the selfgenerated instabilities constrain the solar wind anisotropy
including the thermal core and the suprathermal components. The
generalized power-laws of Kappa-type are the best fitting model for
the observed distributions of particles, and a convenient
mathematical tool for modeling their dynamics. But the anisotropic
Kappa models are not correlated with the observations leading, in
general, to inconsistent effects. This review work aims to reconcile
some of the existing Kappa models with the observations.}

\section{Introduction} 



Direct {\it in-situ} measurements in the solar wind and terrestrial
magnetosphere indicate that the velocity distribution functions
(VDF) of space plasma particles are quasi-Maxwellian up to the mean
thermal velocities (the core component), while they exhibit
non-Maxwellian suprathermal tails (the halo component) at higher
energies (see recent reviews of Marsch (2006)\cite{ma06} and
Pierrard \& Lazar (2010)\cite{pi10}, and references therein).
Processes by which the suprathermal particles are produced and
accelerated \cite{ha85,mi95, ma99, le00, yo06, jo10} are of
increasing interest for applications in astrophysics and laboratory
or fusion plasma devices where they are known as the {\it runaway}
particles decoupled from the thermal state of motion. The solar wind
generally appears to involve an abundance of suprathermal electrons
and ions observed to occur in the interplanetary medium, and their
analysis provides valuable information about their source, whether
it is in the Sun or outer heliosphere.

Accelerated particles (including electrons, protons and minor ions)
are detected at any heliospheric distance in the quiet wind as well
as in the solar energetic particle (SEP) events associated to flares
and coronal mass ejections (CMEs) during solar maximum (see reviews
by Lin (1998)\cite{li98} and Pierrard \& Lazar (2010)\cite{pi10}). A
steady-state suprathermal ion population is observed throughout the
inner heliosphere with a VDF close to $\sim v^{-5}$ \cite{fi06},
and, on the largest scales, the relativistic cosmic-ray gas also
plays such a dynamical role through the galaxy and its halo
\cite{sc02}.

\section{Wave instability constraints of the suprathermal anisotropy}
Nonthermal features and kinetic anisotropies like temperature
anisotropies, heat fluxes or particle streams are also a
characteristic of the solar wind and near the Earth's magnetosphere
\cite{ma06}. At low altitudes in the solar wind, velocity
distributions of plasma particles regularly show an excess of
kinetic energy transverse to the local mean magnetic field ($T_\perp
> T_\parallel$, where $\perp$, and $\parallel$ denote directions
with respect to the magnetic field) most probably due to the
compression exerted by the strong guiding magnetic field near the
Sun. At larger heliospheric distances, the anisotropy is controlled
by the Chew-Goldberger-Low mechanism: the adiabatic expansion of the
poorcollision plasma increases the pressure and temperature along
the magnetic field leading to $T_{\parallel} > T_{\perp}$. In more
violent interplanetary shocks resulting after solar flares and
coronal mass ejections (CME), injection of particle beams into the
ionized interplanetary medium creates additional anisotropy
\cite{ma06,li98}.

\begin{figure}
\sidecaption[t]
\includegraphics[width=5cm,height=6.5cm]{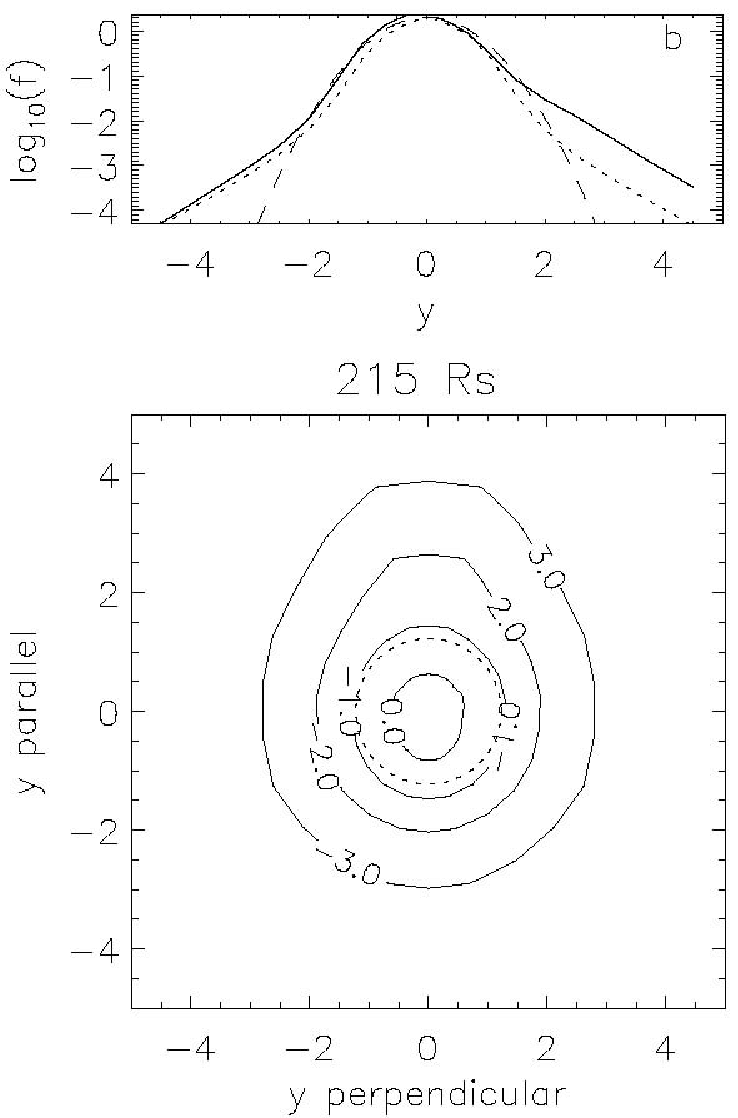}
\caption[width=7cm]{Electron velocity distributions observed by {\it
Wind} mission at 1 AU, as energy spectra (top) parallel (solid) and
perpendicular (dashed) to magnetic field; and velocity space
contours (bottom) in a high-speed solar wind. Note the anisotropic
isodensity contours, less for the core and more pronounced for the
halo and the strahl in fast solar wind (after Pierrard et al. (1999)
\cite{pi99}).} \label{f1}
\end{figure}

The high rate of occurrence of an excess of perpendicular
temperature ($T_{\perp} > T_{\parallel}$) in measurements at large
distances \cite{hel06, st08} is a proof that other mechanisms of
acceleration, namely, the wave-particle interaction, must be at work
there dominating the adiabatic expansion. Indeed, large deviations
from isotropy quickly relax by the resulting wave instabilities,
which act either to scatter particles back to isotropy, or to
accelerate lower energetic particles (Landau or cyclotron heating)
and maintain a suprathermal abundance because thermalization is not
efficient at these scales \cite{ma06}. 

None of these processes is well understood, mostly because these
plasmas are low-collisional and require progress in modeling the
wave turbulence, going beyond MHD models to use a kinetic and
selfconsistent description. In such plasmas transport of matter and
energy is governed by the selfcorrelation between particles and
electromagnetic fields, which can, for instance, convect charged
particles in phase space but are themselves created by these
particles. Thus, the suprathermal populations involve
selfconsistently in both processes of wave turbulence generation and
particle energization, and the resulting Kappa functions that
elegantly describes distributions measured in the solar wind
\cite{pi10}, represent therefore not only a convenient mathematical
tool, but a natural and quite general state of the plasma
\cite{le02}.


\begin{figure}[h] \centering
     \includegraphics[width=5.7cm]{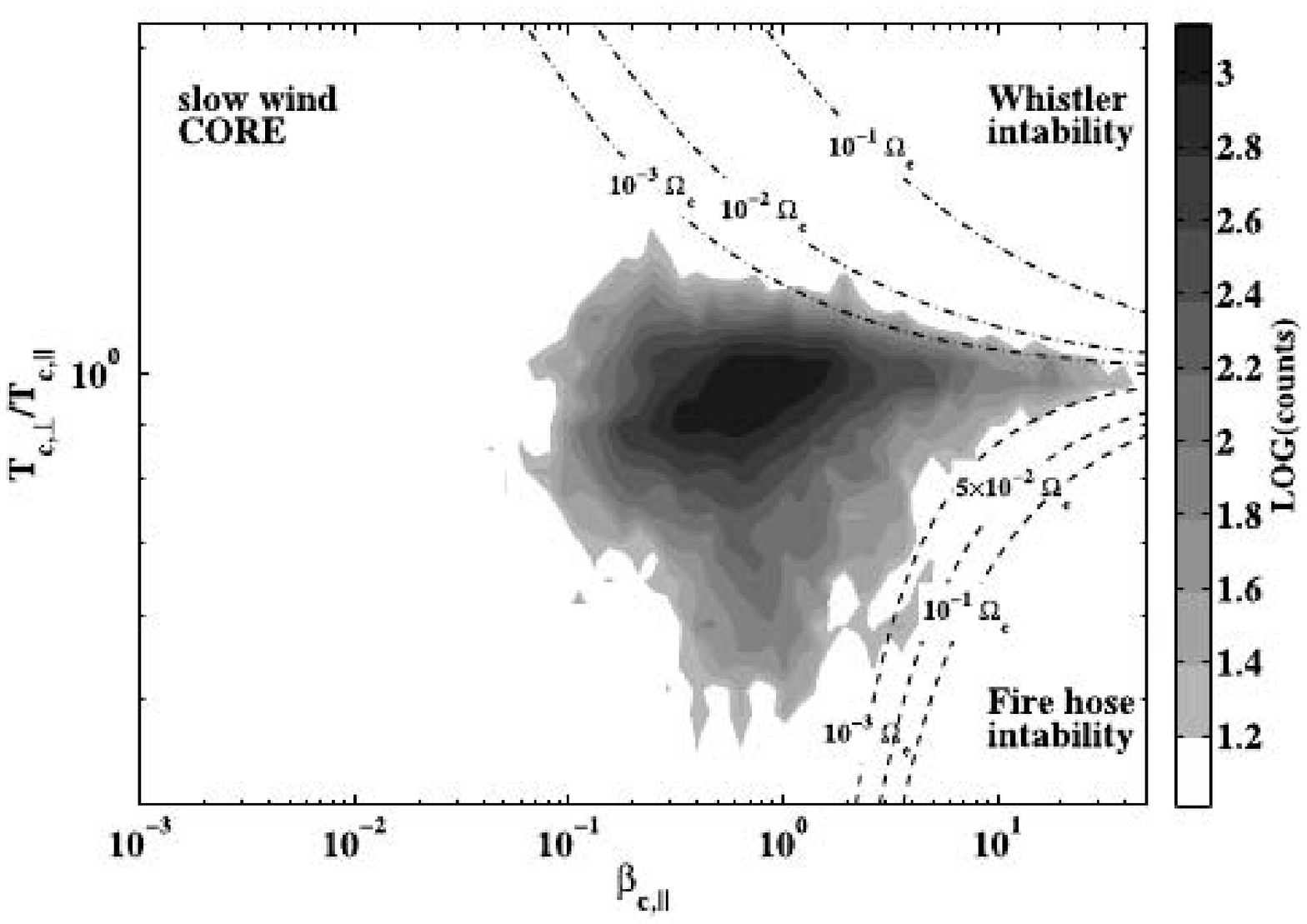} \hspace{0.1cm}
     \includegraphics[width=5.7cm]{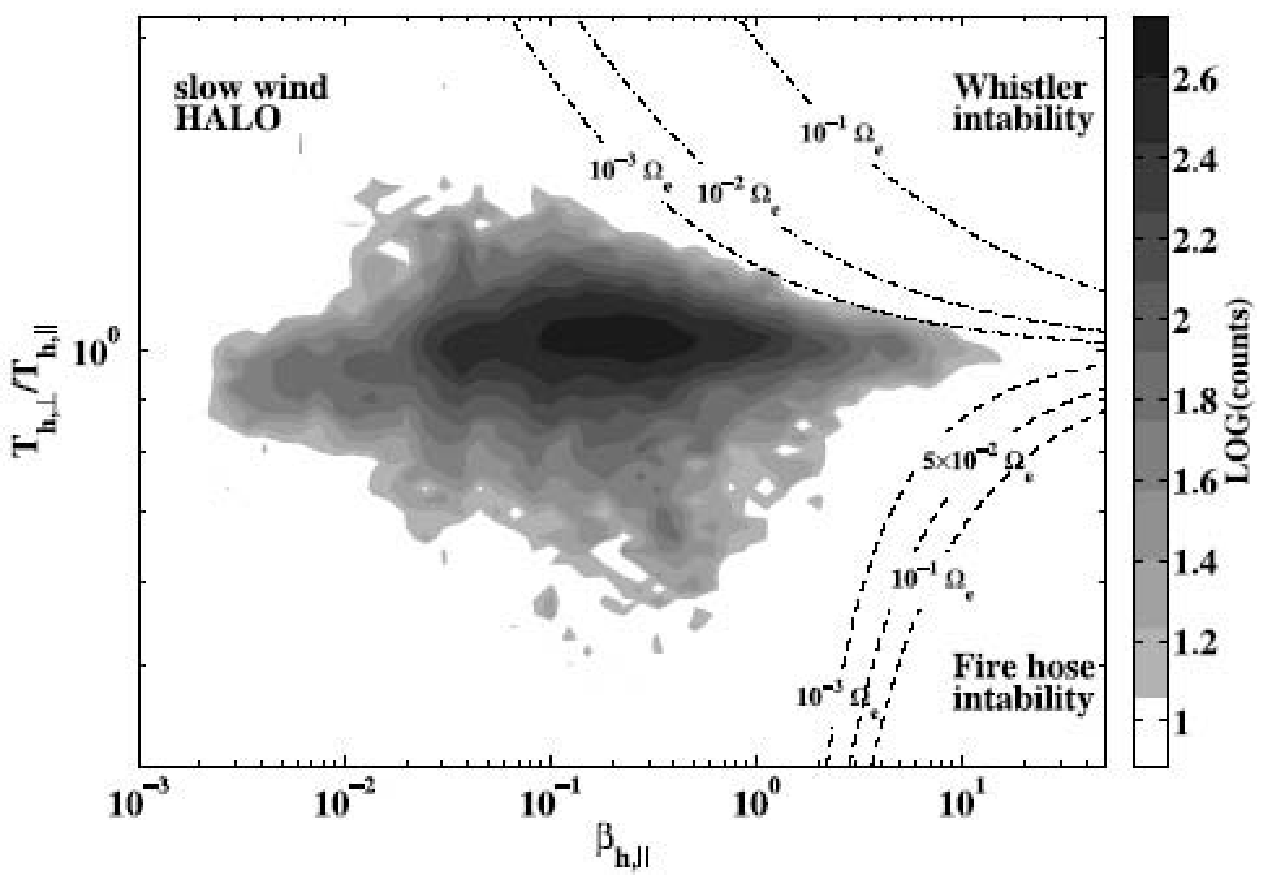}
\caption{Solar wind electron temperature anisotropy
     ($T_{\perp}/T_{\parallel}$) versus $\beta_{\parallel}$: the instability thresholds
     from a bi-Maxwellian model do not constrain the suprathermal halo
     (after Stverak et al. (2008)).} \label{f2}
\end{figure}

While the thermal core in the solar wind is less anisotropic, the
superthermal halo has in general a pronounced temperature anisotropy
($T_{\perp} \ne T_{\parallel}$) with respect to the local magnetic
field \cite{ma06, pi10}. Deformations of the VDFs observed in the
solar wind are not as strong as one would expect from a free motion
of particles (Kasper et al. 2002, Stverak et all. 2008, Bale et al.
2009). Because collisions are not effective, any increase of kinetic
anisotropy is limited by converting the excess of free energy into
electromagnetic fluctuations, and pitch-angle diffusion. Modeling
the VDF with a bi-Maxwellian, the instability thresholds shape
precisely the core (Fig. 2, left) but show regularly an important
departure from to the halo limits (Fig. 2, right). While the
whistler instability limits the perpendicular temperature to grow
($T_{\perp} > T_{\parallel}$), the firehose instability constrains
any excess of parallel temperature ($T_{\perp} < T_{\parallel}$).
Suprathermal particles cannot fit into a Maxwellian approach
($\kappa \to \infty$), but must be modeled with an appropriate Kappa
function. A bi-Kappa (or bi-Lorentzian) function has extensively
been used to model gyrotropic distributions and their
dispersion/stability properties \cite{su91,he05}
\be F_1(v_{\parallel}, v_{\perp}) = {1 \over \pi^{3/2} w_{\perp}^2
w_{\parallel}} \, {\Gamma[\kappa +1] \over \kappa^{3/2}
\Gamma[\kappa -1/2]} \left(1 + {v_{\parallel}^2\over \kappa
w_{\parallel}^2 } + {v_{\perp}^2\over \kappa w_{\perp}^2
}\right)^{-\kappa-1}. \label{e1} \ee
\begin{figure}[h] \centering
     \includegraphics[width=5.7cm, height=4.5cm]{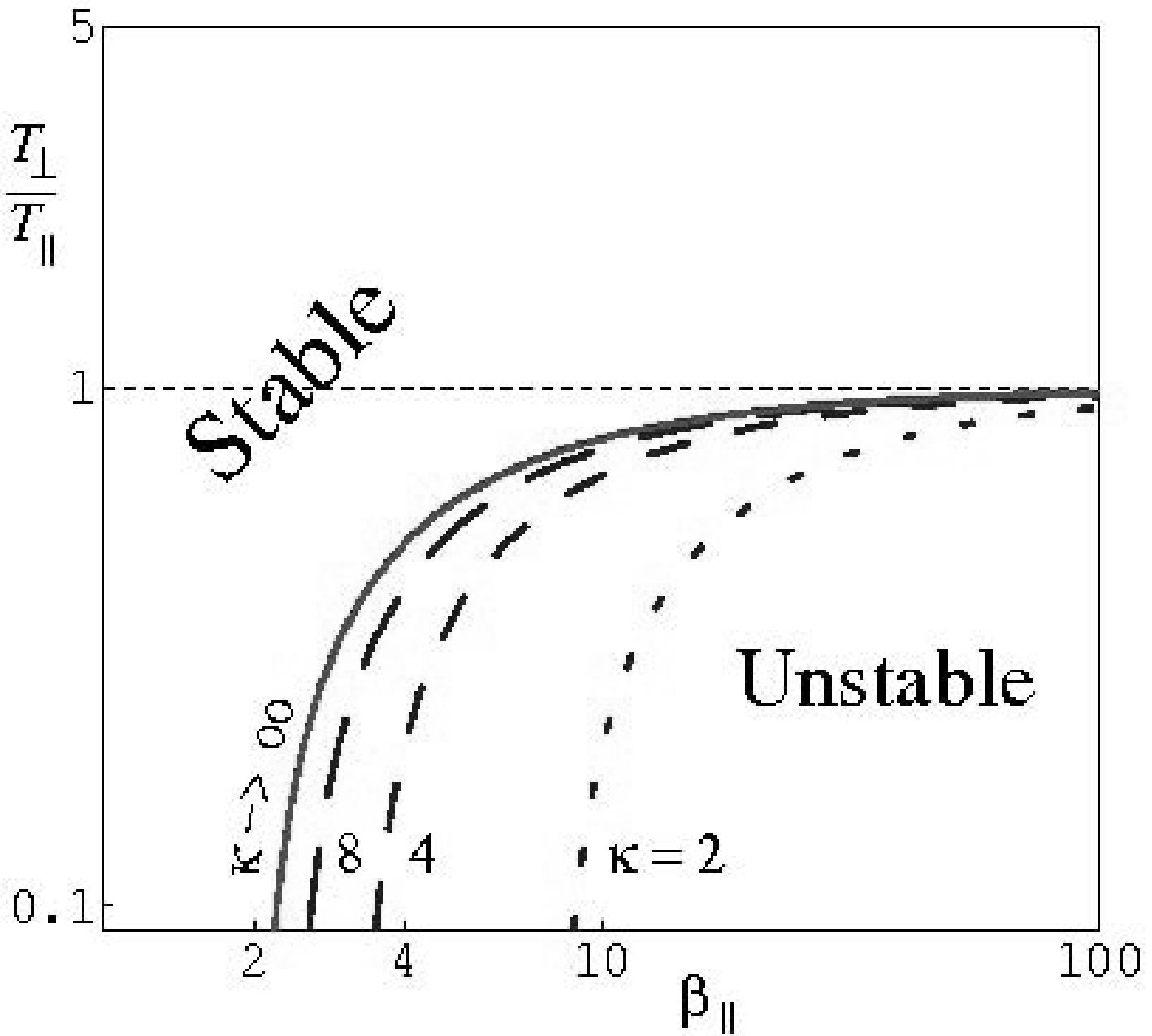} \hspace{0.1cm}
     \includegraphics[width=5.7cm, height=4.5cm]{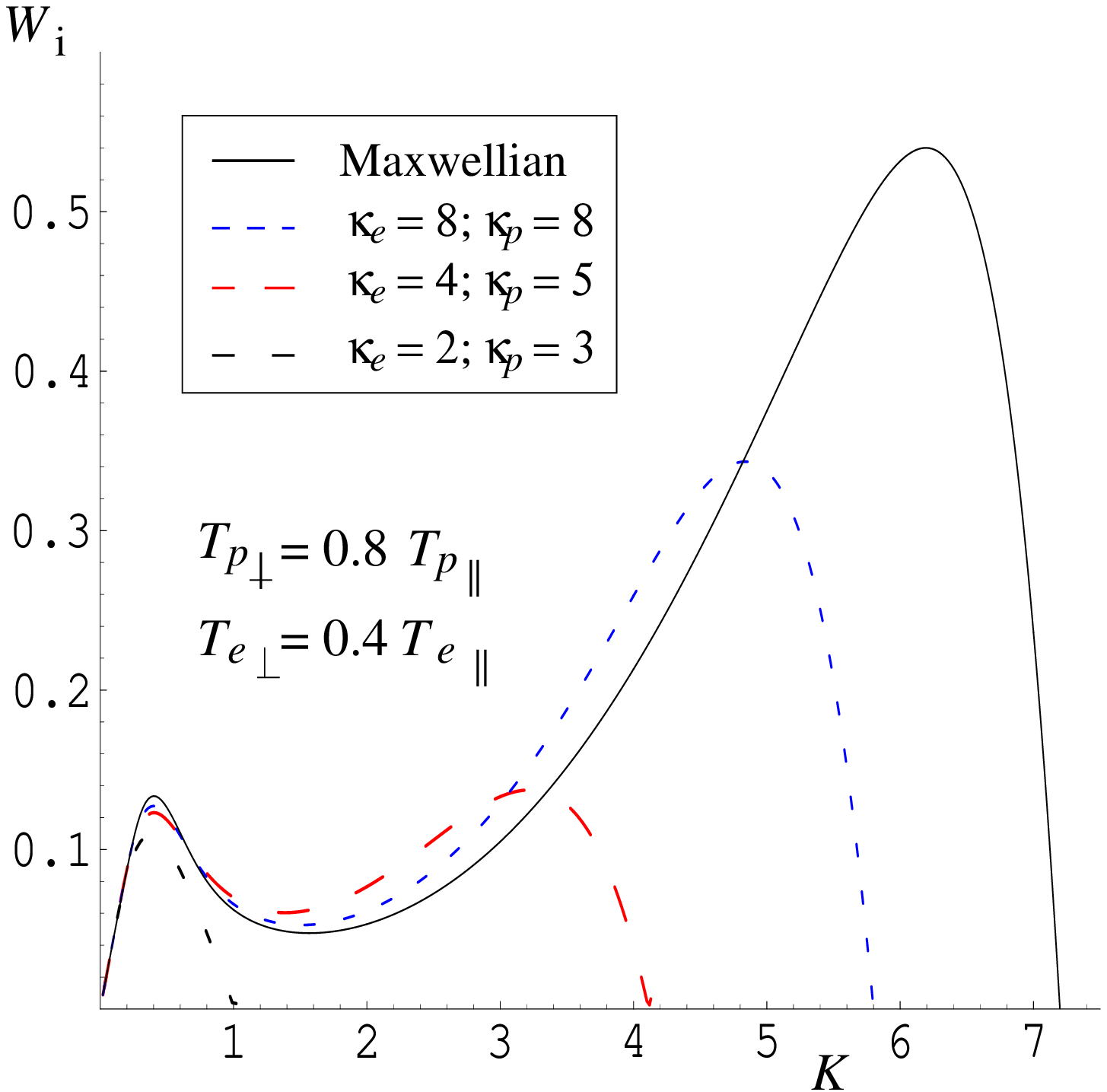}
\caption{Firehose instability: thresholds (left) and growth rates
(right) from a bi-Kappa model.} \label{f3}
\end{figure}
\begin{figure}[h] \centering
     \includegraphics[width=5.2cm, height=3.6cm]{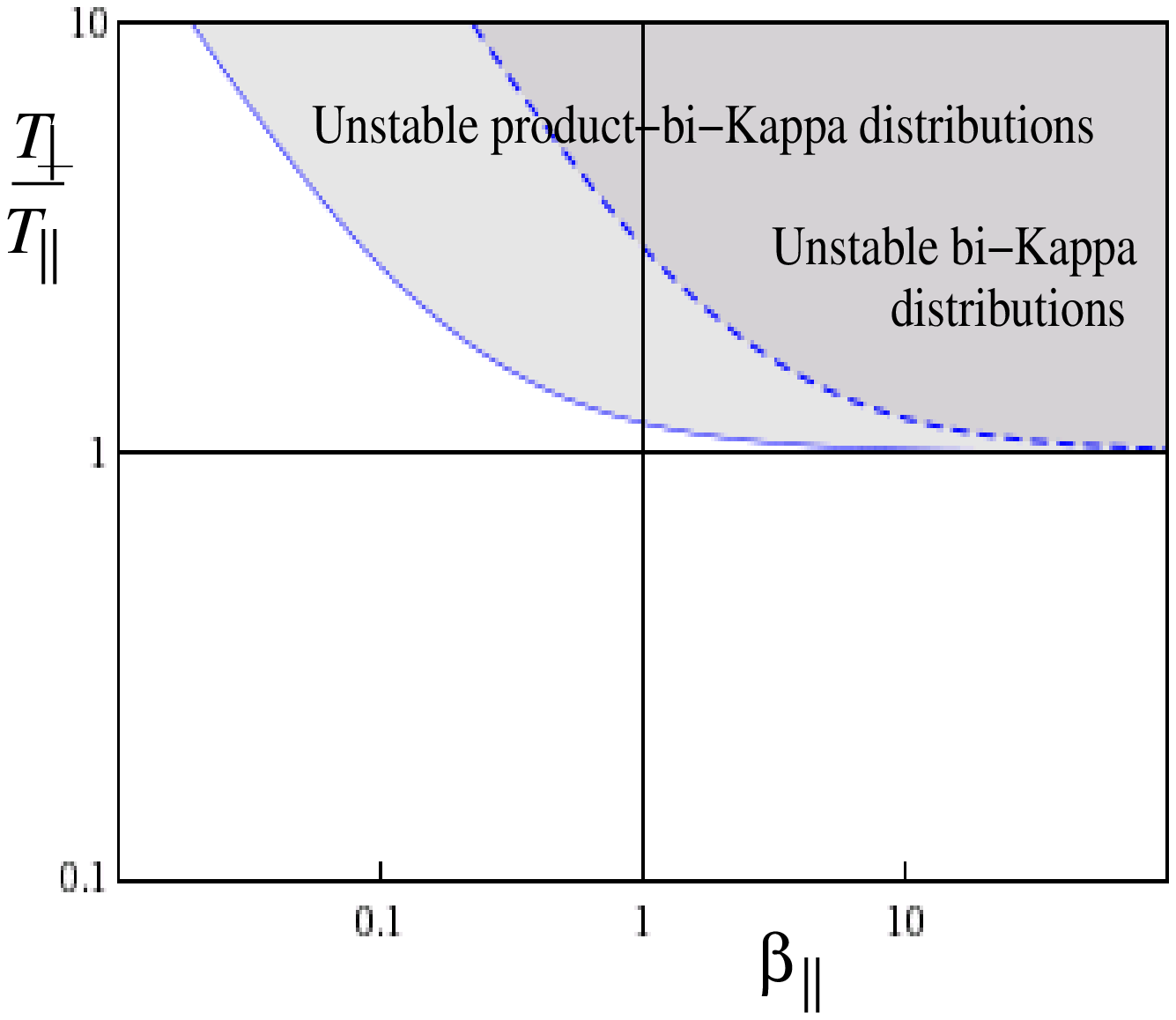} \hspace{0.1cm}
     \includegraphics[width=6cm]{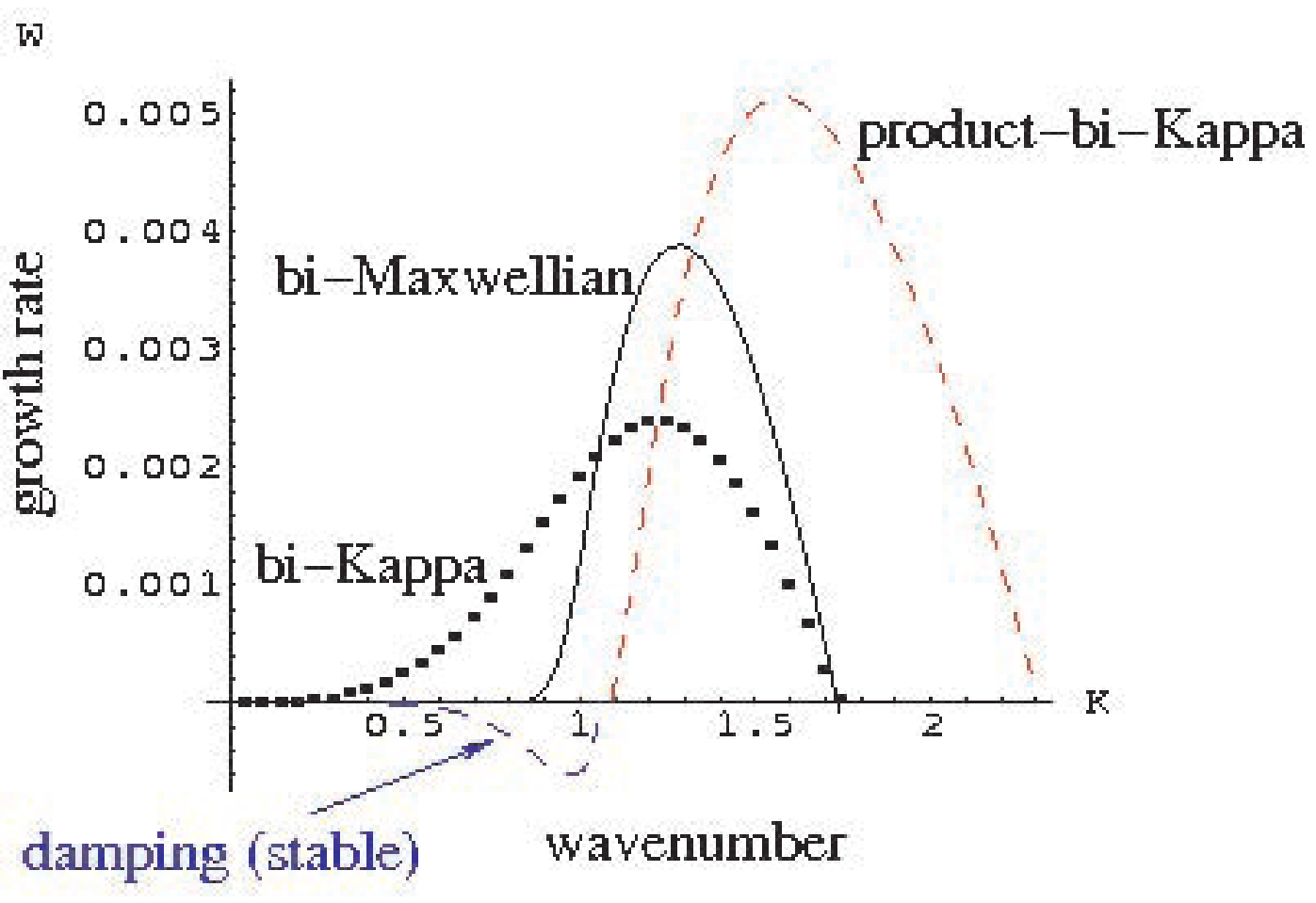}
\caption{Whistler instability: thresholds (left) and growth rates
(right)  from three different models: the bi-Kappa, the
product-bi-Kappa and the bi-Maxwellian (after Lazar et al.
(2011a)).} \label{f4}
\end{figure}

However, this model does not provide the expected better fits to the
observations. For instance, both instabilities of interest for the
electron populations, the whistler and the firehose, are inhibited
and threshold constraints do not approach but rather depart from the
measured limits of the halo \cite{la08,la09,la11a,la11b}. Marginal
conditions of the firehose instability are illustrated in Fig. 3
(left) showing the need for a larger temperature anisotropy and a
larger plasma $\beta_{\parallel} \equiv 8\pi
nk_BT_{\parallel}/B_0^2$ to produce the instability in Kappa
distributed plasmas (low values of $\kappa \to 3/2$). By comparison
to a bi-Maxwellian ($\kappa \to \infty$), the growth rates are in
general reduced and restrained to small wavenumbers (Fig. 3, right)
\cite{la09,la11b}. The same tendency is observed in the evolution of
the whistler instability (Fig. \ref{f4}): by comparison to a
bi-Maxwellian thresholds do not depart much, but the growth rates
decrease significantly for lower values of the index $\kappa$ (Fig.
\ref{f4}, right) \cite{la08,la11a}.

\section{Product-bi-Kappa model}

It is evident that a novel Kappa function is needed to model kinetic
anisotropies and to incorporate the excess of free energy expected
to exist in suprathermal anisotropic plasmas providing better fit to
the observations. Recently a new approach has been proposed
\cite{la10,la11a,bas09} based on the anisotropic product-bi-Kappa
function \cite{su91}
\be F_2 (v_{\parallel}, v_{\perp}) = {w_{\perp}^{-2} \over \pi^{3/2}
w_{\parallel}} \, {\Gamma[\kappa_{\parallel} +1] \over
\kappa_{\parallel}^{1/2} \Gamma[\kappa_{\parallel}+{1 \over 2}]}
\left(1 + {v_{\parallel}^2\over \kappa_{\parallel} w_{\parallel}^2
}\right)^{-\kappa_{\parallel} -1} \left(1 + {v_{\perp}^2\over
\kappa_{\perp} w_{\perp}^2 }\right)^{-\kappa_{\perp}-1}. \label{e2}
\ee
By contrast to the bi-Kappa function (\ref{e1}), which seems to be
less realistic because the two degrees of freedom are coupled and
controlled by the same power index $\kappa$, the new
product-bi-Kappa function shows an advanced flexibility in modeling
gyrotropic VDFs with two distinct temperatures $T_{\parallel,\perp}$
and two distinct power indices $\kappa_{\parallel,\perp}$. Thus, a
new concept for the particle anisotropy can be introduced by
including both anisotropies of the temperatures $T_{\parallel} \ne
T_{\perp}$ and the Kappa indices $\kappa_{\parallel} \ne
\kappa_{\perp}$. The analysis becomes therefore more relevant but
complicated and this is probably the reason this model was only
occasionally invoked (after Summers \& Thorne (1991) proposed it),
merely in a simplified Maxwellian-Kappa form \cite{he05}
\be F_3 (v_{\parallel}, v_{\perp}) = {w_{\perp}^{-2} \over \pi^{3/2}
w_{\parallel}} \, {\Gamma[\kappa +1] \over \kappa^{1/2}
\Gamma[\kappa+1/2]} \left(1 + {v_{\parallel}^2\over \kappa
w_{\parallel}^2 }\right)^{-\kappa -1} \exp \left(- {v_{\perp}^2\over
w_{\perp}^2 }\right). \label{e3} \ee
The distribution function (\ref{e2}) reduces to distribution
function (\ref{e3}) in the limit of very large values of
$\kappa_\perp \to \infty$. Thus both forms of the new model
represented by the general distribution function (\ref{e2}) or by
the particular form (\ref{e3}) reduce to the same bi-Maxwellian in
the limit of very large power indices ($\kappa_{\parallel,\perp} \to
\infty$).
\begin{figure}
\sidecaption[t]
\includegraphics[width=5.5cm]{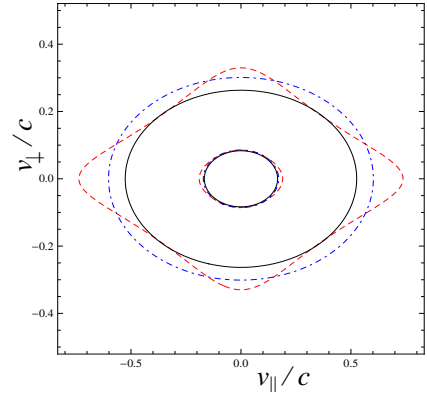}
\caption{Two sets of contours, one close to maximum and one close to
the minimum base: the bi-Maxwellians ($\kappa_\parallel =
\kappa_\perp \to \infty$) with solid lines, product-bi-Kappas with
red dashed lines, and bi-Kappas with blue dotted-dashed lines (for
the same $v_{T,\parallel}/c = 2 v_{T,\perp}/c = 0.2$, and
$\kappa_\parallel = \kappa_\perp = 3$).} \label{f5}
\end{figure}

Contour plots in velocity plane are illustrated in Fig. \ref{f5}
showing no visible excess of asymmetry of the bi-Kappa
(dotted-dashed lines) by comparison to the bi-Maxwellian (solid
lines), but a prominent asymmetry and anisotropy of the new
product-bi-Kappa distribution function (dashed lines) by comparison
to both the bi-Maxwellian and bi-Kappa distribution functions, even
for the same temperatures $T_{\parallel} = T_{\perp}$, and the same
$\kappa_{\parallel} = \kappa_{\perp} = \kappa$ \cite{la10,la11a}.
However, the new distribution model and its dispersion properties
and stability must directly be confronted to the observations in the
solar wind and terrestrial magnetosphere.
\begin{figure}[h] \centering
\includegraphics[width=3.6cm]{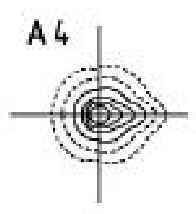} \hspace{0.1cm}
\includegraphics[width=3.6cm]{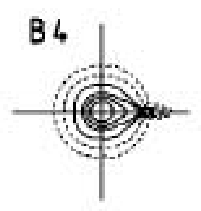} \hspace{0.1cm}
\includegraphics[width=3.6cm]{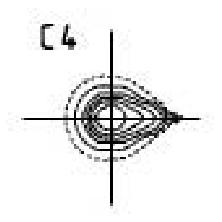}
\caption{The electron VDF with a strahl component in the high speed
solar wind at a distance (in AU) $R=$0.98 (left), $0.64 \le R \le
0.70$ (midle) and $0.29 \le R \le 0.34$ (right) (after Pilipp et al.
1987a).} \label{f6}
\end{figure}

\begin{figure}[h] \centering
\includegraphics[width=3.5cm,height=2.8cm]{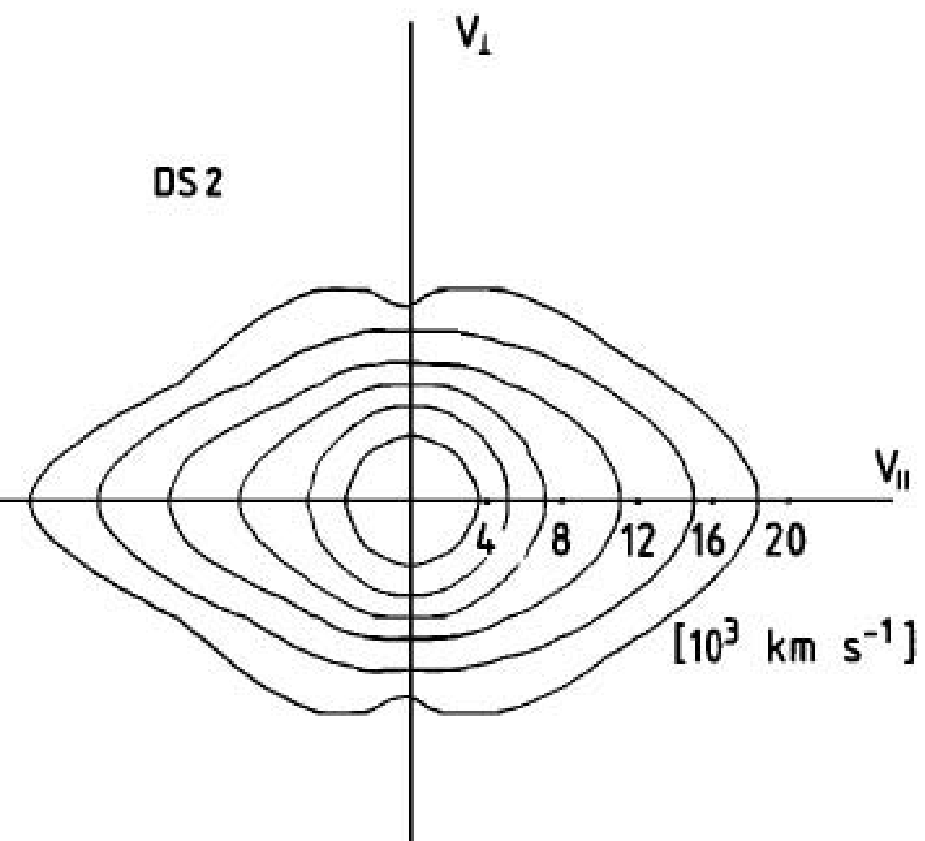}\hspace{0.5cm}
\includegraphics[width=7cm,height=2.8cm]{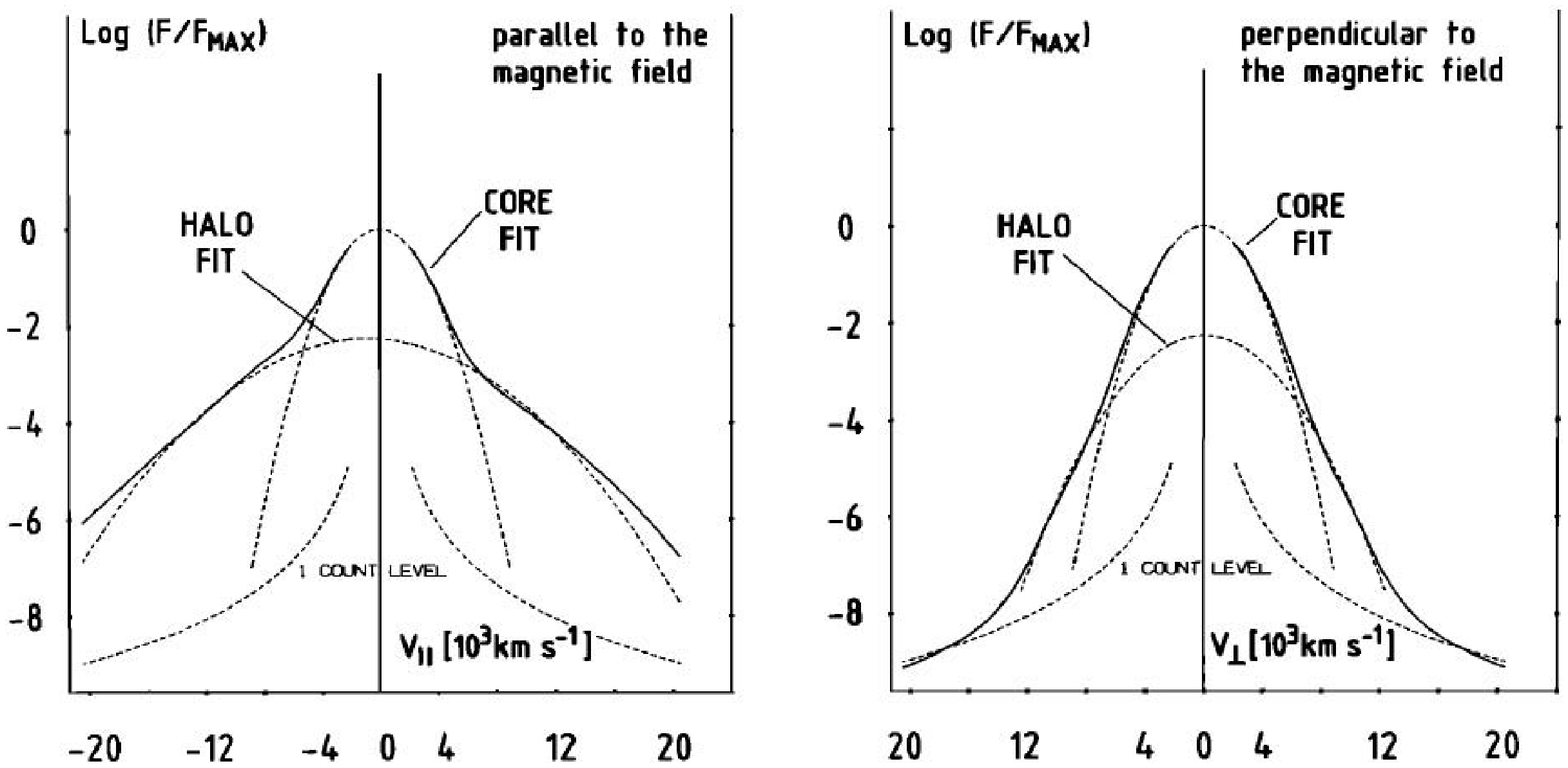}
\caption{Contours and one-dimensional cuts of unusual double-strahl
distribution function observed by Helios (after Pilipp et al.
1987b).} \label{f7}
\end{figure}

Thus, contours of the electron distribution functions measured at
different heliospheric distances in a high speed solar wind (see
Figs. \ref{f6} and \ref{f7}) show such skewed, highly anisotropic
tails along the magnetic field direction  \cite{pi87a,pi87b}, which
look quite similar to the product-bi-Kappa contours in Fig.
\ref{f5}. In Fig. \ref{f6}, skews of the electron distributions are
asymmetric being produced by the magnetic field aligned strahl
population, which is highly energetic (suprathermal) and antisunward
moving \cite{pi87a}. The presumable origin of the strahl electrons
is in the energetic ejecta from the coronal holes, and radial
evolution show a decreasing with distance from the Sun in the favor
of the halo population, which is enhancing \cite{ma05}. The strahl
component is the main driver of the (electron) heat flux, and the
main contributor to the anisotropy of suprathermals, apparently a
manifestation of the adiabatic focusing.

The quite-time distribution in the fast solar wind (Fig. \ref{f6})
is modeled by the product-bi-Kappa function only on one side,
namely, the outward direction ($v_\parallel > 0$), while
semicontours in the backward direction are similar to a bi-Kappa
(low anisotropic) model. Analytically one can combine the Eqs.
(\ref{f2}) and (\ref{f3}) to form
\be F_4 (v_{\parallel}, v_{\perp}) & = & \; H[-v_\parallel] \; C_L
\, \left(1 + {v_{\parallel}^2\over \kappa_L w_{L,\parallel}^2 }+
{v_{\perp}^2\over \kappa_L w_{L,\perp}^2 }\right)^{-\kappa_L-1}
\nonumber
\\ & + & \; H[v_\parallel] \; C_R \; \left(1 + {v_{\parallel}^2\over \kappa_R
w_{R,\parallel}^2 }\right)^{-\kappa_R -1} \exp \left(-
{v_{\perp}^2\over w_{R,\perp}^2 }\right), \label{e4} \ee
where $H[x]$ is the Heaviside function, and $C_{L,R}$ are
normalization constants. Contour plots are similar to the one-sided
strahl observations in Fig. \ref{f6}. Hence the immediate effect
would be a corresponding asymmetric spectrum of the selfgenerated
wave fluctuations, which are expected to dominate the outward
direction. The future work should explore the stability of this new
model and correlate with the observations.

Electron distribution functions (see Fig. \ref{f7}) with an unusual
double strahl have been observed on rare occasions in the solar wind
by Helios probes \cite{pi87b}. Contours plots of these distribution
function are illustrated in Fig. \ref{f7} (left) and show a strong
symmetric bidirectional anisotropy very similar to the
product-bi-Kappa model in Fig. \ref{f5}. Double-strahl distributions
would be observed if the spacecraft was fortuitously near the outer
end of a magnetic loop connected to the Sun, and electrons coming
from both foot points have traveled roughly the same distance and
thus may form a symmetric distribution with a counterstreaming
strahl aspect. A similar distribution is expected in the
observations if the spacecraft was at the outer end of a
disconnected large scale loop \cite{pi87b}.

Another important confirmation should come from the resulting
electromagnetic fluctuations which constrain particle velocity
anisotropy to grow \cite{st08,ka02}. Some preliminary analysis have
indeed shown that the whistler instability growth rates (see Fig.
\ref{f4}, right) calculated for the general product-bi-Kappa model
are enhanced (dashed line) by comparison to those obtained for a
bi-Maxwellian (solid line) or a bi-Kappa model (dotted line), and
the instability thresholds illustrated in Fig. \ref{f4} (left) tend
to approach better the limits of the electron halo in Fig. \ref{f2}
(right) \cite{la11a}. Unlike the bi-Maxwellian or bi-Kappa
distributions, the product-bi-Kappa anisotropy can be stable against
the excitation of the low-wavenumber whistler waves, the critical
wavenumber $ k_c \simeq \Omega/ (\theta_{\parallel}
\sqrt{\kappa_{\parallel} A})$ is vanishing only for
$\kappa_{\parallel} \to \infty$ (Maxwellian plasma). Moreover, it
was shown the instability of an asymmetric Maxwellian-Kappa
distribution without an effective temperature anisotropy ($T_{\perp}
= T_{\parallel}$) against the excitation of the whistler waves
propagating obliquely to the magnetic field \cite{ca07}, most
probably due to the same asymmetry of contours in velocity space
(Fig. \ref{f5}). Thus, the new product-bi-Kappa model seems to
incorporate the excess of free energy expected to exist in
anisotropic suprathermal plasmas. For the opposite case
($T_\parallel > T_\perp$), the firehose instability is currently
under active investigations, and, according to the preceding
discussion, we expect that a product-bi-Kappa model will enhance the
growth rates and provide better agreement of the marginal stability
with the observed limits of the halo (Fig. \ref{f2}, right).


Generation of such asymmetric distribution profiles is also
supported by the anisotropic turbulence of the solar wind, where the
acceleration is expected to occur either perpendicular (cyclotron
damping)\cite{ma99} or in direction (Landau damping)\cite{le00} of
the interplanetary magnetic field. The nonlinear wave-wave and
wave-particle couplings involving intense low-frequency Alfv\'en
waves or electrostatic Langmuir and ion sound (weak) turbulence
driven by the beam-plasma instabilities can also be responsible for
the acceleration in the corona, solar wind and magnetosphere
\cite{mi95,yo06}. However, these models do not provide much guidance
on a full 3D evolution of the distribution function and its
nonthermal features, like suprathermal tails and kinetic
anisotropies, in the acceleration process. This task is still
complicated, but will, hopefully, make the object of the next
investigations.

\section{Conclusions and perspectives}

A large variety of nonthermal features like temperature
anisotropies, heat fluxes or particle streams are permanently
observed in the solar wind and near the Earth's magnetosphere
\cite{ma06,pi10}. These do not grow indefinitely, but tend to
regulate themselves since they contain sufficient free energy to
drive plasma instabilities and micro-turbulence \cite{ba09}. Plasma
wave turbulence provides the main dissipation mechanisms maintaining
a fluid-like behavior of the plasma in spite of the fact that
collisional free paths become comparable to the large scales of the
heliosphere. Bi-Maxwellian models have extensively been used to
describe particle distributions and their dynamics in the solar
wind, but the observations clearly indicate the relevance of Kappa
distribution functions which incorporate both the quasithermal core
and the suprathermal halo. Because the anisotropic Kappa functions
are not correlated, in general, with the observations, here we have
reviewed these models, and established their direct or indirect
confirmations by the observations.

The nondrifting distribution functions used to describe the
temperature anisotropy of the different plasma components (with
respect to the magnetic field), are the bi-Kappa and
product-bi-Kappa functions. The bi-Kappa function seems less
realistic because the two degrees of freedom parallel and
perpendicular to the magnetic field are coupled and controlled by
the same power index k. Instead, the new product-bi-Kappa function
shows more flexibility in modeling the gyrotropic VDFs with two
distinct temperatures $T_{\parallel,\perp} $ and two distinct power
indices $\kappa_{\parallel, \perp}$. A simple comparison of the
contours plots show indeed a significant excess of anisotropy of the
product-bi-Kappa by comparison to both the bi-Kappa and
bi-Maxwellian models.

A further comparison with the contours plots of the measured
distributions indicates two important results of our analysis.
First, the bi-Kappa model is appropriate to describe particle
distributions in the quiet and slow solar wind with a minimum strahl
influence and, in general, small deformations of the halo. Secondly,
the product-bi-Kappa (including the Maxwellian-Kappa) function seems
to be adequate for modeling particle distributions in the fast solar
wind with a prominent strahl component in the direction of magnetic
field. Moreover, a double-strahl distribution looking just like the
product-bi-Kappa contours can be observed near the magnetic large
scale loops with electrons counterstreaming from both foot points.

The selfgenerated wave spectra originating from the Kappa models
exhibit different properties. Thus, the wave instabilities
constrains calculated for a bi-Kappa model do not fit with the halo
limits in a slow solar wind as was expected, but those resulting
from a product-bi-Kappa model seem to shape the halo electrons for
any slow or fast solar wind. This suggests that the new
product-bi-Kappa model incorporates in a proper way the excess of
free energy expected to exist in anisotropic suprathermal plasmas.
Assuming that the same wave fluctuations are the most plausible
mechanism of acceleration and formation of suprathermal populations,
we conclude pointing out the relevance of our study by the fact that
an appropriate Kappa model enables a powerful selfconsistent
analysis as it should be itself a product of particle acceleration
by the selfexcited fluctuations.

\begin{acknowledgement}
The authors acknowledge financial support from the Research
Foundation Flanders (project G.0729.11), the KU Leuven (project
GOA/2009-009, grant F/07/061), ESA Prodex~10 (project C~90205) and
by the Deutsche Forschungsgemeinschaft (DFG), grant Schl 201/21-1.
Financial support by the European Commission through the SOLAIRE
Network (MTRN-CT-2006-035484), and the Seventh Framework Program
(FP7/2007-2013) the grant agreement SWIFF (project nr. 2633430,
www.swiff.eu) is gratefully acknowledged.
\end{acknowledgement}

\end{document}